 \documentclass[final,5p,times]{elsarticle}

 \usepackage{graphics}

\usepackage{amssymb}

\journal{NIM A  RICAP-2013}

\begin{document}

\begin{frontmatter}



\title{ Scientific verification of High Altitude Water Cherenkov observatory}


\author[1]{Antonio Marinelli}
\author[2]{Kathryne Sparks}
\author[1]{Ruben Alfaro}
\author[3]{Mar\'{i}a Magdalena Gonz\'{a}lez}
\author[3]{Barbara Patricelli}
\author[3]{Nissim Fraija}

\author[]{\\for the HAWC collaboration}
\address[1]{Instituto de F\'{i}sica, Universidad Nacional Aut\'{o}noma de M\'{e}xico, Mexico D.F., Mexico}
\address[2]{Department of Physics, Pennsylvania State University, University Park, PA, USA}
\address[3]{Instituto de Astronomia, Universidad Nacional Aut\'{o}noma de M\'{e}xico, Mexico D.F., Mexico}
\begin{abstract}
The High Altitude Water Cherenkov (HAWC) observatory is a TeV gamma-ray and cosmic-ray detector currently under construction at an altitude of 4100 $m$ close to volcano
Sierra Negra in the state of Puebla, Mexico. The HAWC \cite{HAWCgrb} observatory is an extensive air-shower array comprised of 300 optically-isolated water Cherenkov detectors (WCDs).
Each WCD contains $\sim$200,000 liters of filtered water and four upward-facing photomultiplier tubes. In Fall 2014, when the HAWC observatory will reach an area of 22,000
$m^{2}$, the sensitivity will be 15 times higher than its predecessor Milagro \cite{milagro2006}. Since September 2012, more than 30 WCDs have been instrumented and taking data. This first
commissioning phase has been crucial for the verification of the data acquisition and event reconstruction algorithms. Moreover, with the increasing number of instrumented
WCDs, it is important to verify the data taken with different configuration geometries. In this work we present a comparison between Monte Carlo simulation and data recorded by the experiment 
during 24 hours of live time between 14 and 15 April of 2013 when 29 WCDs were active.
\end{abstract}

\begin{keyword}
\emph{Water Cherenkov observatory} \sep \emph{scientific verification} \sep \emph{Monte Carlo simulations}


\end{keyword}

\end{frontmatter}


\section{Introduction}
\begin{figure}[!htb]
\centering
\includegraphics[width=0.5\textwidth]{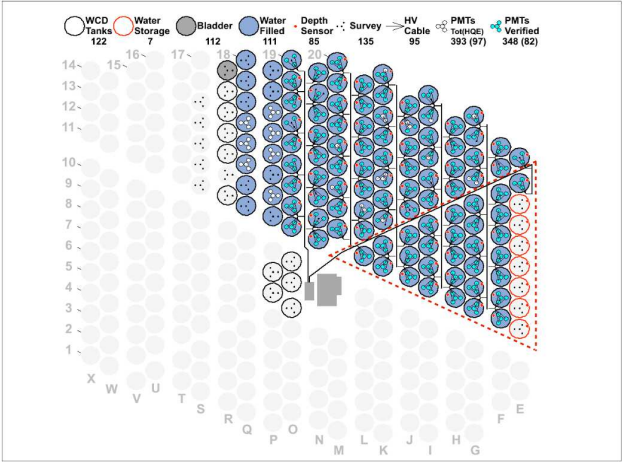}
\caption{The actual status of HAWC construction. The red dashed line distinguishes the HAWC-30 phase validated with Monte Carlo in this work.}\label{HAWCconf}
\end{figure}

The construction of the High Altitude Water Cherenkov observatory \cite{HAWCgrb} started in January 2012 and will be finished in the fall of 2014. 
Two intermediate commissioning phases have been selected  by the
collaboration:  the first one with 30 operative water Cherenkov detectors (WCDs) and the second one with 100 operative WCDs. The first
commissioning phase was completed in September 2012 with the 30 WCD configuration shown in Fig. \ref{HAWCconf}. The scientific verification of
this first phase was very important to test the hardware and software components of the experiment and give a first validation of the
observation capabilities of HAWC. A fundamental step of the scientific verification is the comparison of data with Monte Carlo simulations. This
comparison verifies the reconstruction algorithms for cosmic ray shower events and helps optimize the calibrations of the HAWC photomultipliers
tubes (PMTs). In this work we compare to Monte Carlo the multiplicity of reconstructed events, the obtained zenith and azimuth angle distributions
and the X-Y reconstructed shower core distribution.
\section{HAWC observatory}
HAWC is an array under construction at Pico de Orizaba national park (Mexico) by a joint US-Mexican collaboration. The main reason for the construction of this observatory is to obtain a survey of Very High Energy (VHE) gamma-ray emitters in the portion of the sky between $-20^{\circ}$ and $+60^{\circ}$ of declination. The large field of view and the continuous data taking will make possible this goal in the energy range between 100 GeV and 100 TeV. Most of the expertise for building the detector is obtained from the Milagro observatory \cite{milagro2006}, operated between 2000 and 2008 at an altitude of 2630 m in the Jemez mountains, near Los Alamos (New Mexico). Milagro was designed to trigger on air shower components at ground level using a large pool of water with 723 PMTs instrumented in an optically isolated reservoir. The HAWC collaboration utilizes part of the electronics and the 8-inch PMTs from Milagro while the geometry was redesigned in order to get a detector with a sensitivity 15 times higher. The final configuration of HAWC will be composed of 300 WCDs. Each WCD is comprised of a steel tank 7.3 m in diameter and 4.5 m in height, a plastic bladder containing 200 000 liters of purified water, and four PMTs: three 8-inch Hamamatsu R5912 PMTs and one 10-inch R7081-MOD high-quantum efficiency PMT. The gain in sensitivity \cite{HAWCSensi} compared to the Milagro experiment is reached thanks to the higher elevation with respect to sea level, the large area covered by the array, the greater optical isolation of the PMTs and the introduction of the new 10-inch PMTs. The main data acquisition (DAQ) system of
the experiment uses a time to digital converter (TDC) to process the signal coming from the front-end board (FEB) \cite{Milagrito} and discriminate between leading and trailing edges with two discrimination thresholds ($\sim1/4$ photoelectron, $\sim5$ photoelectrons). This technique accurately measures each pulse width (or ToT) that can be used to obtain the charge of the pulse over a large
dynamic range. The secondary DAQ (scaler) system collects information from all the PMTs triggering on only the low threshold of $\sim1/4$ of a photoelectron and is not able to provide directional information. This analysis only uses the data obtained with the primary DAQ and processed with the online software trigger.
\section{Monte Carlo simulations}
The simulation chain used by the HAWC experiment proceeds through four main code modules: CORSIKA, HAWCSim, AERIE and HMC-Analysis.
The complete Monte Carlo chain traces every particle produced by the interaction of primary cosmic ray in the atmosphere down to its Cherenkov
radiation cutoff energy, and each Cherenkov photon is tracked within the detector volume. CORSIKA \cite{corsika92} is used to generate and simulate
air shower events produced by gamma-ray primaries and eight different cosmic-ray nuclei. The output of this stage produces files with particle content
saved at the HAWC altitude. Events are not simulated with their natural spectrum but are given a relatively flat spectrum to over sample the rare high-energy events. This bias is later removed by the application of proper weights obtained with HMC-Analysis. 
HAWCSim is a Monte Carlo detector simulation code, based on the GEANT4 \cite{geant4} toolkit, that takes the particle information from CORSIKA (which propagates the cosmic ray production up to 10 m above the detector) and simulates events up to the point where they produce photoelectrons (PEs) in each PMT. The output of this step is an ascii file with a list of PEs and water hits. 
The next step in the Monte Carlo analysis chain, AERIE, simulates the collection of PEs by the HAWC PMTs and the DAQ hardware. 
Using the charge and timing of triggered PMT hits \cite{HAWCSensi} we obtain the cores and the directions of the reconstructed cosmic ray events. All the events that would trigger the detector and a prescaled selection of events that fail to trigger the detector are written in ROOT \cite{root} files. At this step the raw hit information is removed and a summary record of each event remains. The prescaled selection of non-triggering events is included in the summary information to preserve how many events were thrown during the simulation for effective area calculations. The output of this stage is one ROOT file for each simulated species. With HMC-Analysis, we take the nine species input files from the reconstruction step and assign weights to the events based on their species, their core distance and their energy. The purpose of this Monte Carlo step is to take the raw unphysical spectrum, simulated core distances, and composition and give each event a weight to obtain a realistic Monte Carlo production of one second of physical data.
\section{Validation of HAWC-30 data with Monte Carlo simulation}
\begin{figure}[!htb]
\centering
\includegraphics[width=0.5\textwidth]{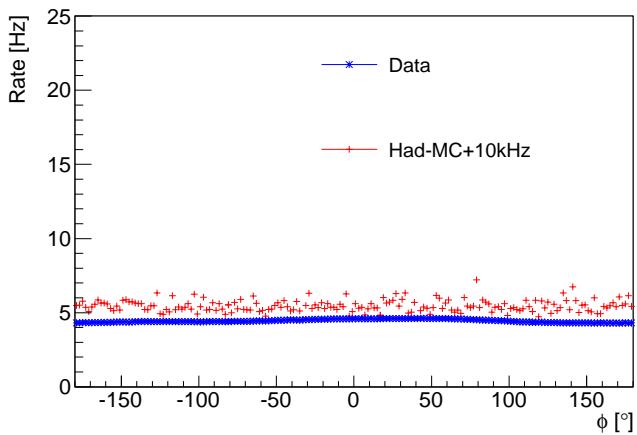}
\caption{Azimuth angle distribution of reconstructed events. The blue stars show the 
data while the red crosses represent the Monte Carlo expectations. The cuts applied to Monte Carlo and 
data are the same as used for the previous plots.}\label{HAWCAzimuth}
\end{figure}

\begin{figure}[!htb]
\centering
\includegraphics[width=0.5\textwidth]{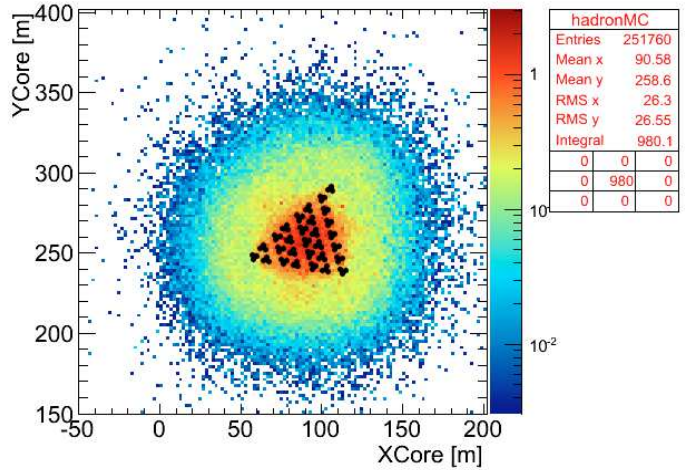}
\caption{The bidimensional distribution of the reconstructed core of the events generated with Monte Carlo. The black dots represent the position of the PMTs for the simulated configuration of the detector: 29 WCDs and 116 PMTs. The color scale represents the frequency of events, given in Hz. This plot is for all the primary hadronic components simulated by CORSIKA.}\label{HAWCCore-MC}
\end{figure}

As mentioned previously, the validation of data taken with the 30 WCD phase with Monte Carlo simulation is a fundamental step of the scientific verification of the HAWC observatory. The Monte Carlo sample used in this analysis is composed of the weighted reconstructed events from a homogeneous  cosmic ray production with a spectral index of 2.7 and a random noise of 10 kHz applied to each PMT. The complete Monte Carlo chain described in Section 3 
has been used to obtain the reconstructed shower events. The data sample used is obtained from $\sim$24 hours of TDC reconstructed data taken between 14 and 15 of April 2013. 
The time of reconstructed events has been corrected by a time calibration and the complete set of data has been normalized to obtain the equivalent experiment life time of one second (like the Monte Carlo production). Considering the portion of time when data taking was offline for instrumentation of the detector and for software development, we apply to the reconstructed data a dead time of $5\%$. For this analysis data and Monte Carlo have been filtered with some cuts. In particular we have eliminated all the events reconstructed with less than 30 TDC channels and all the events that are reconstructed with an angle greater than $60^{\circ}$ from the zenith of the detector. These cuts are applied to reject the events that are poorly reconstructed and avoid problems with the limits of the reconstruction algorithm and the angular acceptance of the PMTs.   
\begin{figure}[!htb]
\centering
\includegraphics[width=0.5\textwidth]{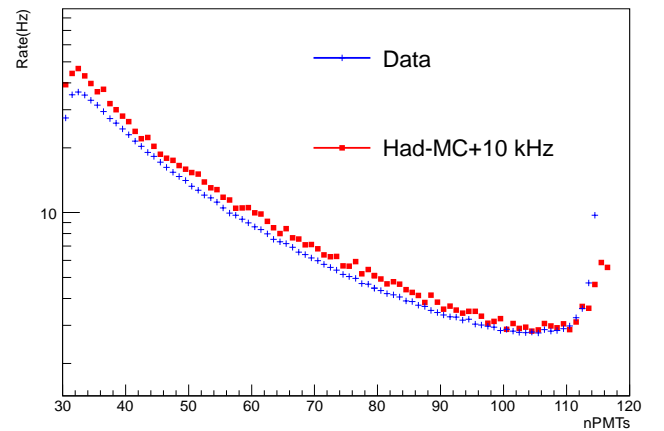}
\caption{Comparison between data and Monte Carlo for multiplicity distribution of reconstructed events. The red squares show the Monte Carlo distribution while the blue cross shows the data distribution.
The same cuts have been applied to Monte Carlo and data: we excluded events reconstructed with less than 30 channels and an angle greater than $60^\circ$ from the zenith of the detector. Data was corrected for the  $5\%$ dead time of the detector. The ratio between the integrals of the two distributions is about 0.85.}\label{HAWCMult}
\end{figure}
As we can see in Fig. \ref{HAWCMult}, the multiplicity distribution of reconstructed events obtained from the data is in accordance with the multiplicity obtained from the Monte Carlo simulation. The difference of $15\%$ between the integrals of the two distributions is acceptable considering the difference between Monte Carlo and data. In particular the Monte Carlo events are weighted with the results of CREAM-2 \cite{Cream} experiment (the uncertainty of these data and the difference between the two experiments are possible causes) and the effect due to the temperature and pressure variation at the HAWC site are not taken into account. 
\begin{figure}[!htb]
\centering
\includegraphics[width=0.5\textwidth]{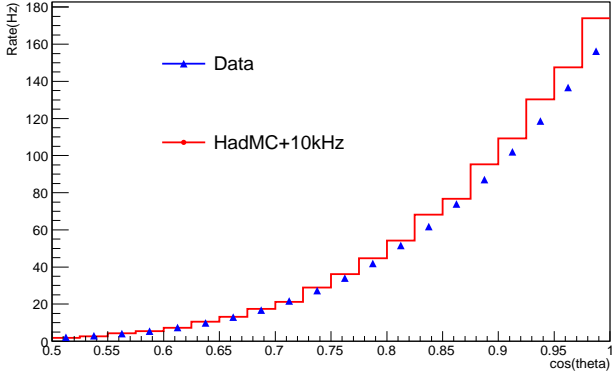}
\caption{Zenith angle distribution of reconstructed events, with the blue triangles representing the real data while the red histogram shows the Monte Carlo expectation. The cuts applied to data and Monte Carlo simulation are the same explained before. The data has been corrected for the dead time of $5\%$ of the detector}\label{HAWCZenith}
\end{figure}
Figures \ref{HAWCZenith} and \ref{HAWCAzimuth}, showning the comparison between Monte Carlo and data for the zenith and azimuth angle distributions of events, give us proof of a correct reconstruction of data. The $15\%$ of excess of Monte Carlo reconstructed events is distributed in the cone with a aperture of $25^{\circ}$ from the zenith of the detector.\\ 
\begin{figure}[!htb]
\centering
\includegraphics[width=0.5\textwidth]{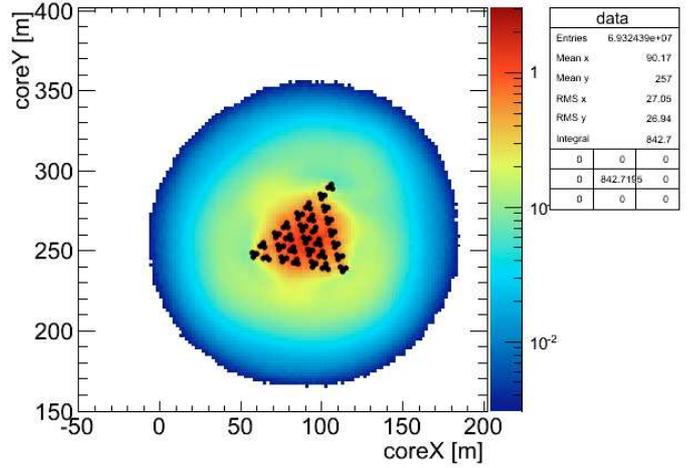}   
\caption{The bidimensional (X and Y) distribution of the reconstructed core of the events from 
data. The black dots represent the position of the PMTs
that took data on April 14-15. During the 24 hours of this sample of data, 29 WCDs and 114 PMTs were active. The color scale represents the frequency of events, expressed in Hz. The cuts applied to the data are the same as used for the other plots.}\label{HAWCCore-data}
\end{figure}
Another important validation of correct reconstruction of 
data for this first commissioning phase has been done through the comparison of the cores of the shower events. From the comparison of Fig. 
\ref{HAWCCore-data} and Fig. \ref{HAWCCore-MC}, we can conclude that the shape and the size of the core of reconstructed events from 
data is very well reproduced by the core distribution obtained with the simulated reconstructed events.
\section{Conclusions}
This work shows the validation of data from the first commissioning phase of the HAWC observatory with a Monte Carlo simulation. This construction phase
was important for testing the DAQs and the offline reconstruction algorithms.
In addition to the observation of VHE emission from astrophysical sources, the Monte Carlo validation of  
data is considered a fundamental step of the HAWC scientific verification. The variables used for the comparison with Monte Carlo simulations
are the reconstructed event multiplicity, the zenith and azimuth angle distributions and the X-Y reconstructed shower core distribution. This comparison shows that, for all the variables with the cuts considered, the
data are in agreement with the Monte Carlo simulation up to a difference of $15\%$. We suppose that this difference can be due to the uncertainty in the weights used for Monte Carlo events and to the absence in the simulation of the environmental variability effects.   
\section*{Acknowledgments}
We thank Brian Baughman and Jim Braun for their great help with this analysis. 
We gratefully acknowledge Scott DeLay his dedicated efforts in the construction
and maintenance of the HAWC experiment. This work has been supported
by: the National Science Foundation, the US Department of Energy Office
of High-Energy Physics, the LDRD program of Los Alamos National Laboratory,
Consejo Nacional de Ciencia y Tecnologia (grants 55155, 103520, 105033,
105666, 122331 and 132197), Red de Fisica de Altas Energias, DGAPA-UNAM
(grants IN105211, IN112910 and IN121309, IN115409, IN108713), VIEP-BUAP (grant 161-
EXC-2011), the University of Wisconsin Alumni Research Foundation, and the
Institute of Geophysics and Planetary Physics at Los Alamos National Lab.



\bibliographystyle{elsarticle-num}
\bibliography{/Users/AntonioMarinelli/HAWC/RICAP13/elsarticleRicap13/biblio}

\begin{thebibliography}{1}
\expandafter\ifx\csname url\endcsname\relax
  \def\url#1{\texttt{#1}}\fi
\expandafter\ifx\csname urlprefix\endcsname\relax\def\urlprefix{URL }\fi
\expandafter\ifx\csname href\endcsname\relax
  \def\href#1#2{#2} \def\path#1{#1}\fi

\bibitem{HAWCgrb}
{Abeysekara A.~U. et al.}, {On the sensitivity of the HAWC observatory to
  gamma-ray bursts}, Astroparticle Physics 35 (2012) 641--650.
\newblock \href {http://arxiv.org/abs/1108.6034} {\path{arXiv:1108.6034}},
  \href {http://dx.doi.org/10.1016/j.astropartphys.2012.02.001}
  {\path{doi:10.1016/j.astropartphys.2012.02.001}}.

\bibitem{milagro2006}
{MILAGRO Collaboration}, {Recent Results from the Milagro Gamma Ray
  Observatory}, Nuclear Physics B Proceedings Supplements 151 (2006) 101--107.
\newblock \href {http://dx.doi.org/10.1016/j.nuclphysbps.2005.07.018}
  {\path{doi:10.1016/j.nuclphysbps.2005.07.018}}.

\bibitem{HAWCSensi}
{Abeysekara A.~U. et al.}, {Sensitivity of the High Altitude Water Cherenkov
  Detector to Sources of Multi-TeV Gamma Rays}, ArXiv e-prints\href
  {http://arxiv.org/abs/1306.5800} {\path{arXiv:1306.5800}}.

\bibitem{Milagrito}
{R. Atkins and et al.}, Milagrito, a tev air-shower array, Nuclear Instruments
  and Methods in Physics Research Section A: Accelerators, Spectrometers,
  Detectors and Associated Equipment 449~(3) (2000) 478 -- 499.
\newblock \href {http://dx.doi.org/10.1016/S0168-9002(00)00146-7}
  {\path{doi:10.1016/S0168-9002(00)00146-7}}.

\bibitem{corsika92}
J.~N. {Capdevielle}, P.~{Grieder}, J.~{Knapp}, P.~{Gabriel}, H.~J. {Gils},
  D.~{Heck}, H.~J. {Mayer}, J.~{Oehlschl{\"a}ger}, H.~{Rebel}, G.~{Schatz},
  T.~{Thouw}, {The Karlsruhe extensive air shower simulation code CORSIKA.},
  1992.

\bibitem{geant4}
{S. Agostinelli et al.}, Geant4Ña simulation toolkit, Nuclear Instruments and
  Methods in Physics Research Section A: Accelerators, Spectrometers, Detectors
  and Associated Equipment 506~(3) (2003) 250 -- 303.
\newblock \href
  {http://dx.doi.org/http://dx.doi.org/10.1016/S0168-9002(03)01368-8}
  {\path{doi:http://dx.doi.org/10.1016/S0168-9002(03)01368-8}}.

\bibitem{root}
R.~{Brun}, F.~{Rademakers}, {ROOT - An object oriented data analysis
  framework}, Nuclear Instruments and Methods in Physics Research A 389 (1997)
  81--86.
\newblock \href {http://dx.doi.org/10.1016/S0168-9002(97)00048-X}
  {\path{doi:10.1016/S0168-9002(97)00048-X}}.

\bibitem{Cream}
{Maestro, P. et al.}, {Elemental energy spectra of cosmic rays measured by
  CREAM-II}, ArXiv e-prints\href {http://arxiv.org/abs/1003.5759}
  {\path{arXiv:1003.5759}}.

\end{thebibliography}






\end{document}